\RequirePackage{lineno}
\documentclass[twocolumn,secnumarabic,amssymb,nobibnotes,aps,prd]{revtex4}
\usepackage{amssymb, amsmath}
\usepackage{mathbbold}
\usepackage{graphicx}
\usepackage{dcolumn}
\usepackage{bm}
\usepackage{hyperref}
\usepackage{color}
\begin{document}

\title{ Functional Renormalization for Chiral and $U_A(1)$ Symmetries at Finite Temperature}
\author{Yin Jiang}
\author{Pengfei Zhuang}
\affiliation{Physics Department, Tsinghua University, Beijing
100084, China }
\date{\today}

\begin{abstract}
We investigated the chiral symmetry and $U_A(1)$ anomaly at finite
temperature by applying the functional renormalization group to the
$SU(3)$ linear sigma model. Expanding the local potential around the
classical fields, we derived the flow equations for the
renormalization parameters. In chiral limit, the flow equation for
the chiral condensate is decoupled from the others and can be
analytically solved. The Goldstone theorem is guaranteed in vacuum
and at finite temperature, and the two phase transitions for the
chiral and $U_A(1)$ symmetry restoration happen at the same critical
temperature. In general case with explicit chiral symmetry breaking,
the two symmetries are partially and slowly restored, and the scalar
and pseudoscalar meson masses are controlled by the restoration in
the limit of high temperature.
\end{abstract}

 \maketitle

\section{Introduction}
\label{s1} The chiral symmetry is one of the fundamental properties
of quantum chromodynamics (QCD). In chiral limit with zero current
quark mass, the QCD Lagrangian density respects the symmetry of
$U_L(3)\times U_R(3)=U_V(1)\times U_A(1)\times SU_V(3)\times
SU_A(3)$ at classical level. In vacuum, the condensate of right-hand
quark and left-hand anti-quark $\langle \bar q_L q_R\rangle$ breaks
the $SU_A(3)$ symmetry, and the $U_A(1)$ is broken by the anomaly
due to the nontrivial topology of principal bundle of gauge
field\cite{thooft}. The chiral symmetry breaking leads to a rich
meson and baryon spectrum, and as a supplement the $U_A(1)$ anomaly
explains the non-degeneracy of $\eta$ and $\eta'$ mesons\cite{eta}.
As a strong interacting system should approach its classic limit at
high temperature, the chiral symmetry is believed to be restored in
hot medium. The relation between the two symmetries in vacuum and at
finite temperature has been studied for a long time. While the two
symmetries are expected to be restored\cite{u1restore} at high
temperature, it is still an open question whether the two phase
transitions happen under the same condition.

The lattice simulation is a powerful tool to study the QCD
symmetries in vacuum and at finite temperature. The chiral
condensate is observed to decrease with increasing temperature and
the chiral susceptibility shows a peak at a critical temperature
$T_c$. By a proper definition, the topological charge and its
susceptibility are used to describe the $U_A(1)$ anomaly in the pure
gauge field theory and the unquenched
theory\cite{lattice1,lattice2}. In both cases, the susceptibility
drops down above the critical temperature $T_c$ of the chiral
restoration and the charge keeps an obvious deviation from zero at
high temperature $T>T_c$. The simulation for the instanton model
shows such a partial restoration too\cite{iilm}. On the other hand,
the observation of the hadron spectra provides an experimental way
to test the restoration of the two symmetries in hot medium created
in relativistic heavy ion collisions. The mass shift due to the
chiral restoration enhances or reduces the hadron thermal
production, for instance for the kaon yields and
ratios\cite{kaon1,kaon2,ratio}. The partial restoration of the
$U_A(1)$ symmetry is closely related to the production of $\eta$
meson and spin-excited hadrons in hot medium\cite{u1eta,u1hydro}.

The effective models\cite{bubblelsm,u1lsm} are often used to
describe the QCD phase structure, especially at finite baryon
density where the lattice simulation meets the fermion sign problem
and can not yet present precise result. For the study of chiral
symmetry at low energy, the effective models at hadron level can
include only scalar and pseudoscalar mesons. In this work we choose
the linear sigma model which has been widely discussed in
vacuum\cite{chiral}. The simplest version of this model is with the
symmetry of $SU_L(2)\times SU_R(2)$ or $O(4)$, including only the
Goldstone modes $\pi$ and their partner, the $\sigma$ meson. The
linear sigma model is considered as a good laboratory for various
approximation methods like mean field and the
Cornwall-Jackiw-Tomboulis (CJT) approximation. In the frame of this
model, the $\sigma$ and $\pi$ properties have been deeply
investigated in vacuum and hot medium\cite{mfsu2lsm,su2cjt}. The
model is also extended to include fundamental represented quarks
coupled to the mesons in Yukawa form. However, in the $SU(2)$
version the $U_A(1)$ anomaly can not be properly studied, since
there are no enough flavors for $\eta$ and $\eta'$ mesons which are
related to the anomaly. In order to study the chiral symmetry and
$U_A(1)$ symmetry at the same time, the flavor symmetry group is
chosen as $SU_L(3)\times SU_R(3)$ in this work.

The pion, kaon and eta mesons are the Goldstone modes corresponding
to the spontaneous $SU_A(3)$ symmetry breaking. These zero mass
mesons are guaranteed by the Nambu-Goldstone theorem in vacuum as
well as at low temperature $T<T_c$. However, in the mean filed
approximation\cite{mfsu2lsm} and the CJT
approximation\cite{cjt,su2cjt}, the Nambu-Goldstone theorem is
destroyed at low temperature. In this work we use the functional
renormalization group (FRG) method to study the chiral symmetry and
$U_A(1)$ symmetry in the $SU(3)$ linear sigma model at finite
temperature. As a non-perturbative
method\cite{frgreview1,frgreview2,frgreview3}, the FRG has been used to study
phase transitions in various systems like cold atom
gas\cite{frgcoldatom}, nucleon gas\cite{frgnucleon} and hadron
gas\cite{frgpqmm,o4frg,onfrgmeson1,onfrgmeson2,su2frg,su2frg2}. By
solving the flow equation which connects physics at different
momentum scales, the FRG shows a great power to describe the phase
transitions and the corresponding critical phenomena, which are
normally difficult to be controlled in the mean field approximation
because of the absence of the quantum fluctuations. Instead of
adding hot loops to the thermodynamic potential in usual ways of
going beyond mean field, the FRG effective potential at tree level
of mean field approximation includes already quantum fluctuations
through the mass and coupling constant renormalization and can
guarantee the Nambu-Goldstone theorem in the chiral symmetry
breaking phase.

We proceed as follows. In Section \ref{s2}, we briefly review the
$SU(3)$ linear sigma model, following the notation in \cite{cjt},
and apply the functional renormalization to the model. The flow
equations are derived in general case with explicit chiral breaking
terms, and the one for the chiral condensate can be solved
analytically in chiral limit. In Section \ref{s3} we present the
numerical results for the light and strange quark condensates and
the topological susceptibility at finite temperature. Finally in
Section \ref{s4} we summarize our results.

\section{ Application of Functional Renormalization to the $SU(3)$ Linear Sigma Model}
\label{s2}
The $SU_L(3)\times SU_R(3)$ linear sigma model has been
widely studied in mean field and CJT approximation in vacuum and at
finite temperature\cite{cjt,su3lsm,su3oneloop}. Following the
notations in \cite{cjt}, the Lagrangian density of the model is
expressed as
\begin{eqnarray}
\mathcal{L}&=&\text{Tr}\left(\partial_\mu\phi^\dagger\partial^\mu\phi-m^2\phi^\dagger\phi\right)
+c\left[\text {Det}(\phi)+\text {Det}(\phi^\dagger)\right]\nonumber\\
&-&\lambda_1\left[\text {Tr}(\phi^\dagger\phi)\right]^2
-\lambda_2\text {Tr}(\phi^\dagger\phi)^2+\text
{Tr}\left[H(\phi+\phi^\dagger)\right],\nonumber\\
\label{lang}
\end{eqnarray}
where the meson matrix $\phi=T_a\phi_a$ and the trace Tr are defined
in the flavor space, the meson field $\phi_a=\sigma_a+i\pi_a$
contains the scalar part $\sigma_a$ and the pseudoscalar part
$\pi_a$, the $3\times 3$ Gell-Mann matrices $T_a=\lambda_a/2$ for
$a=1,\cdots, 8$ and $T_0=\sqrt{2/3}$ for $a=0$ obey the relations
$\text {Tr}(T_aT_b)=\delta_{ab}/2,\ \left[T_a,
T_b\right]=if_{abc}T_c$ and $\left\{T_a, T_b\right\}=d_{abc}T_c$
with the structure constants $f_{abc}$ and $d_{abc}$, $m^2$ is the
mass parameter, and $c$, $\lambda_1$ and $\lambda_2$ are the
coupling constants.

The Lagrangian density (\ref{lang}) is invariant under the
$SU_L(3)\times SU_R(3)$ transformation, except the last term which
explicitly breaks the chiral symmetry,
\begin{equation}
\text {Tr}\left[H(\phi+\phi^\dagger)\right]=h_a\sigma_a,
\label{break}
\end{equation}
where the matrix $H$ is defined as $H=h_a T_a$ with 9 parameters
$h_a$.

The determinant term in (\ref{lang}) explicitly breaks the $U_A(1)$
symmetry which in QCD is violated by the anomaly. If the coefficient
$c$ of the $U_A(1)$ anomaly term vanishes, the symmetry group of the
system is enlarged to $U_L(3)\times U_R(3)$.

In vacuum and at finite temperature but zero density, there are only
scalar condensates
\begin{equation}
\langle\phi\rangle=T_a\langle\sigma\rangle_a,
\label{conden}
\end{equation}
where $\langle X\rangle$ means the ensemble average of the operator
$X$. To simplify the notation, we use $\bar\sigma$ to replace
$\langle\sigma\rangle$ in the following. Making a shift for the
meson field $\phi\to\langle\phi\rangle+\delta\phi$ and substituting
it into the Lagrangian density (\ref{lang}), the effective potential
of the system at classical level can be written as
\begin{eqnarray}
U(\bar\sigma)&=&\frac{m^2}{2}\bar\sigma^2_a
-G_{abc}\bar\sigma_a\bar\sigma_b\bar\sigma_c
+\frac{1}{3}F_{abcd}\bar\sigma_a\bar\sigma_b\bar\sigma_c\bar\sigma_d\nonumber\\
&-&h_a\bar\sigma_a,
\label{u1}
\end{eqnarray}
and the dynamical masses generated by the condensates can be
extracted from the coefficients of the term $(\delta\phi)^2$ and
form two $9\times 9$ matrices $M_S$ and $M_P$ for the scalar and
pseudoscalar mesons,
\begin{eqnarray}
(M^2_S)_{ab}&=&m^2\delta_{ab}-6G_{abc}\bar{\sigma}_c+4F_{abcd}\bar{\sigma}_c\bar{\sigma}_d,\nonumber\\
(M^2_P)_{ab}&=&m^2\delta_{ab}+6G_{abc}\bar{\sigma}_c+4H_{abcd}\bar{\sigma}_c\bar{\sigma}_d
\label{mass1}
\end{eqnarray}
with the coefficients defined as
\begin{eqnarray}
G_{abc}&=&\frac{c}{6}\Big[d_{abc}+\frac{9}{2}d_{000}\delta_{a0}\delta_{b0}\delta_{c0}\nonumber\\
&-&\frac{3}{2}\left(\delta_{a0}d_{0bc}+\delta_{b0}d_{a0c}+\delta_{c0}d_{ab0}\right)\Big],\nonumber\\
F_{abcd}&=&\frac{\lambda_1}{4}\left(\delta_{ab}\delta_{cd}+\delta_{ad}\delta_{bc}+\delta_{ac}\delta_{bd}\right)\nonumber\\
&+&\frac{\lambda_2}{8}\left(d_{abe}d_{ecd}+d_{ade}d_{ebc}+d_{ace}d_{ebd}\right),\nonumber\\
H_{abcd}&=&\frac{\lambda_1}{4}\delta_{ab}\delta_{cd}\nonumber\\
&+&\frac{\lambda_2}{8}\left(d_{abe}d_{ecd}+f_{ade}d_{ebc}+f_{ace}d_{ebd}\right).
\label{coe}
\end{eqnarray}

The physical condensates are determined by minimizing the potential,
\begin{equation}
\frac{\partial U(\bar\sigma)}{\partial \bar\sigma_a}=0
\end{equation}
which leads to the gap equations
\begin{equation}
m^2\bar{\sigma}_a-3G_{abc}\bar{\sigma}_b\bar{\sigma}_c
+\frac{4}{3}F_{abcd}\bar{\sigma}_b\bar{\sigma}_c\bar{\sigma}_d-h_a=0.
\label{gap1}
\end{equation}

In different thermodynamic environments, the condensate
$\langle\phi\rangle$ can be further simplified. Since the isospin
zero mesons $\sigma_0$ and $\sigma_8$ or $\sigma_\eta$ and
$\sigma_{\eta'}$ can couple to the vacuum without violating Lorentz
invariance and parity, the classical field matrix
$\langle\phi\rangle$ and the coefficient matrix $H$ in the chiral
breaking term contain only two components,
$\langle\phi\rangle=T_0\bar\sigma_0+T_8\bar\sigma_8$ and
$H=T_0h_0+T_8h_8$. In order to simplify the expressions, we make a
rotation for the condensates $\bar\sigma_0$ and $\bar\sigma_8$ and
the chiral breaking parameters $h_0$ and $h_8$,
\begin{eqnarray}
\bar\sigma_u &=& \sqrt{\frac{2}{3}}\bar\sigma_0+\sqrt{\frac{1}{3}}\bar\sigma_8,\nonumber\\
\bar\sigma_s &=&
\sqrt{\frac{1}{3}}\bar\sigma_0-\sqrt{\frac{2}{3}}\bar\sigma_8,\nonumber\\
h_u &=& \sqrt{\frac{2}{3}}h_0+\sqrt{\frac{1}{3}}h_8,\nonumber\\
h_s &=& \sqrt{\frac{1}{3}}h_0-\sqrt{\frac{2}{3}}h_8.
\label{rotate}
\end{eqnarray}

In terms of the rotated condensates $\bar\sigma_u$ and
$\bar\sigma_s$ and the rotated breaking parameters $h_u$ and $h_s$,
the classical potential and the gap equations can be explicitly
expressed as
\begin{eqnarray}
U(\bar\sigma_u,\bar\sigma_s)&=&\frac{m^2}{2}\left(\bar\sigma^2_u+\bar\sigma^2_s\right)
-\frac{c}{2\sqrt{2}}\bar\sigma^2_u\bar\sigma_s\nonumber\\
&+&\frac{\lambda_1 }{4} \left(\bar\sigma^2_u+\bar\sigma^2_s\right)^2
+\frac{\lambda_2}{8} \left(\bar\sigma^4_u+2\bar\sigma^4_s\right),
\label{u2}
\end{eqnarray}
and
\begin{eqnarray}
&&
m^2\bar\sigma_u-\frac{1}{\sqrt{2}}c\bar\sigma_u\bar\sigma_s+\lambda_1\left(\bar\sigma^2_u+\bar\sigma^2_s\right)\bar\sigma_u
+\frac{1}{2}\lambda_2\bar\sigma^3_u=h_u,\nonumber\\
&& m^2\bar\sigma_s-\frac{1}{2\sqrt{2}}c\bar\sigma^2_u+\lambda_1
\left(\bar\sigma^2_u+\bar\sigma^2_s\right)\bar\sigma_s+\lambda_2\bar\sigma^3_s=h_s.
\label{gap2}
\end{eqnarray}
For the meson mass matrices $M_S$ and $M_P$, each has only one
independent off-diagonal element $M^2_{08}=M^2_{80}$ and four
independent diagonal elements $M^2_{00}$, $M^2_{88}$ and
\begin{eqnarray}
\label{mass2}
&& m^2_{a_0}=(M^2_S)_{11}=(M^2_S)_{22}=(M^2_S)_{33},\\
&& m^2_{\kappa}=(M^2_S)_{44}=(M^2_S)_{55}=(M^2_S)_{66}=(M^2_S)_{77},\nonumber\\
&& m^2_{\pi}=(M^2_P)_{11}=(M^2_P)_{22}=(M^2_P)_{33},\nonumber\\
&&
m^2_K=(M^2_P)_{44}=(M^2_P)_{55}=(M^2_P)_{66}=(M^2_P)_{77},\nonumber
\end{eqnarray}
and diagonalizing the meson subspace $a=0,8$ generates the
pseudoscalar mesons $\eta$ and $\eta'$ and the corresponding scalar
mesons. There are six parameters in the model, the mass $m$, the
three coupling constants $c, \lambda_1$ and $\lambda_2$ and the two
chiral breaking parameters $h_u$ and $h_s$. They should be
determined by the experimental data in vacuum. Firstly, the partial
conservation of axial-vector current (PCAC) leads to a relation
between the condensates and the pion and kaon decay constants
$f_\pi$ and $f_K$,
\begin{eqnarray}
\bar\sigma_u &=& f_\pi,\nonumber\\
\bar\sigma_s &=& \frac{-f_\pi+2f_K}{\sqrt{2}},
\label{pcac}
\end{eqnarray}
then the gap equations (\ref{gap2}) can be reexpressed by the
Goldstone modes $\pi$ and $K$,
\begin{eqnarray}
h_u &=& m^2_\pi f_\pi,\nonumber\\
h_s &=& {-m^2_\pi f_\pi+5m^2_Kf_K\over \sqrt 2},
\label{h}
\end{eqnarray}
and the combination of the isospin zero pseudoscalar mesons $\eta$
and $\eta'$ determines the couplings $\lambda_2$ and $c$,
\begin{eqnarray}
\lambda_2 &=& 2\frac{\sqrt 6\bar\sigma_sm^2_K-\sqrt
2\bar\sigma_0m^2_\pi+\bar\sigma_8(m^2_\eta+m^2_{\eta'})}{(\bar\sigma^2_u+4\bar\sigma^2_s)\sigma_8},\nonumber\\
c &=& \frac{m^2_K-m^2_\pi}{f_K-f_\pi}-\lambda_2(2f_K-f_\pi).
\end{eqnarray}
Substituting the above obtained parameters into any of the two gap
equations, one can get the relation between the mass $m$ and the
coupling $\lambda_1$. By fitting a scalar meson mass, for instance
$m_\sigma$, one can then separately fix the two parameters. In
summary, the six parameters $m$, $c$, $\lambda_1$, $\lambda_2$,
$h_u$ and $h_s$ are fitted by the experimental values of $m_\pi$,
$m_K$, $f_\pi$, $f_K$, and $m^2_\eta+m^2_{\eta'}$ and one of the
scalar meson masses. It is necessary to note that, in this way one
can fix a group of parameters, but the obtained condensates may not
correspond to the minimum of the potential. One should check the
secondary derivative of the potential, $\partial^2
U/\partial\bar\sigma_i\partial\bar\sigma_j >0$.

We now apply the functional renormalization group to the $SU(3)$
linear sigma model. The core quantity in the frame of FRG is the
averaged effective action $\Gamma_k$ at a momentum scale $k$ in
Euclidean space,
\begin{equation}
\Gamma_k[\langle\phi\rangle]=\int {d^4x } \left[\text
{Tr}\left(Z_{k}\partial_\mu\langle\phi\rangle^\dagger\partial^\mu\langle\phi\rangle\right)+U_k(\langle\phi\rangle)+\cdots\right],
\label{gamma1}
\end{equation}
where $Z_k$ is the wave function renormalization constant,
$U_k(\langle\phi\rangle)$ is the classical potential (\ref{u2}) but
with renormalized mass and coupling parameters $m_k, c_k,
\lambda_{1k}$ and $\lambda_{2k}$ and scale dependent condensates
$\bar\sigma_{uk}$ and $\bar\sigma_{sk}$, and the symbol $\cdots$
stands for the high order terms of the field $\langle\phi\rangle$.
The scale dependence of the averaged action is characterized by the
flow equation\cite{frgreview1,frgreview2,frgreview3} in momentum
representation,
\begin{equation}
\frac{\partial\Gamma_k[\langle\phi\rangle]}{\partial
k}=\frac{1}{2}\int{d^4 p\over (2\pi)^4} \text {
Tr}\left[\left(\Gamma^{(2)}_k[\langle\phi\rangle]+R_k\right)^{-1}\frac{\partial
R_k}{\partial k}\right], \label{flow1}
\end{equation}
where $\Gamma_k^{(2)}$ is the second order functional derivative of
the averaged action
$\Gamma^{(2)}_k[\langle\phi\rangle]=\delta^2\Gamma_k/\delta\langle\phi\rangle^2$,
and the infrared cutoff function $R_k$ which is used to suppress the
quantum fluctuations at low momentum $p<k$ is chosen as the
optimized regulator function $R_k=(k^2-p^2)\theta(k^2-p^2)$
\cite{litim}. From our numerical calculation shown in the next
section, the symmetry restoration and meson mass spectra at finite
temperature are not sensitive to the choice of the cutoff
function\cite{regulator}.

Following the effective action flow starting from the ultraviolet momentum
$k=\Lambda $, the physics we are interested in could be obtained at
$k=0$.

By assuming the space-time independence of the classical field
$\langle\phi\rangle$, the effective action to the lowest order is
determined by the classical potential only,
\begin{equation}
\Gamma_k=\int {d^4x } U_k\left(\langle\phi\rangle\right).
\label{gamma2}
\end{equation}
With the known infrared cutoff function $R_k$, after doing the
three-momentum integration for the mesons at finite temperature, the
FRG flow equation can be simplified as
\begin{eqnarray}
\label{flow2}
\partial_k U_k&=& \frac{Z_{Sk}^{-1}k^4}{6\pi^2}\left(1-{\eta_{Sk}\over
6}\right)T\sum_n \text {Tr}
D_{Sk}\nonumber\\
&+& \frac{Z_{Pk}^{-1}k^4}{6\pi^2}\left(1-{\eta_{Pk}\over
6}\right)T\sum_n\text {Tr} D_{Pk}
\end{eqnarray}
with the meson propagators $D_{Sk}^{-1}=
Z_{Sk}^{-1}\left(\omega^2_n+k^2\right)+M_S^2$ and  $D_{Pk}^{-1}=
Z_{Pk}^{-1}\left(\omega^2_n+k^2\right)+M_P^2$ and the definition
$\eta_k=-k\partial_k Z_k^{-1}/Z_k^{-1}$, where $\omega_n=2n\pi T$
with $n=0,1,2 \cdots$ is the meson Matsubara frequency in the
imaginary time formalism of finite temperature field theory, and we
have considered different renormalization constants $Z_{Sk}$ and
$Z_{Pk}$ for the scalar and pseudoscalar fields. In order to
complete the set of flow equations, we need equations for the
evolution of $Z_{Sk}$ and $Z_{Pk}$\cite{frgreview3}. To the one-loop
level they read
\begin{eqnarray}
\label{zk} -\partial_k Z_{Sk}^{-1}&=&
\frac{Z_{Sk}^{-2}k^4}{6\pi^2}T\sum_n\text {Tr}
\left(D_{Sk}^2\Gamma_{SSS}D_{Sk}^2\Gamma_{SSS}\right)\\
&+& \frac{Z_{Pk}^{-2}k^4}{6\pi^2}T\sum_n\text {Tr}
\left(D_{Pk}^2\Gamma_{SPP}D_{Pk}^2\Gamma_{SPP}\right),\nonumber\\
-\partial_k Z_{Pk}^{-1}&=&
\frac{Z_{Sk}^{-1}Z_{Pk}^{-1}k^4}{3\pi^2}T\sum_n\text {Tr}
\left(D_{Pk}^2\Gamma_{PSP}D_{Sk}^2\Gamma_{PSP}\right),\nonumber
\end{eqnarray}
where
$\left(\Gamma_{SSS}\right)^a_{bc}=\partial^3U_k/\partial\bar\sigma_a\partial\bar\sigma_b\partial\bar\sigma_c,\
\left(\Gamma_{SPP}\right)^a_{bc}=\partial^3U_k/\partial\bar\sigma_a\partial\bar\pi_b\partial\bar\pi_c$
and
$\left(\Gamma_{PSP}\right)^a_{bc}=\partial^3U_k/\partial\bar\pi_a\partial\bar\sigma_b\partial\bar\pi_c$
are $9\times 9$ matrices for the three-line vertexes with a fixed
external meson $a$. In the above discussion we have assumed that the
wave function renormalization constant depends only on the Lorentz
transformation property of the mesons, but is independent of the
detailed meson types. In the following numerical calculations
related to the three-line vertexes, we take the Goldstone mode
$a=4$.

As we will see from the numerical calculation in the next section,
the contribution from the wave function renormalization to the
thermodynamics of the system is very small and can be safely
neglected as a first order approximation. In this case, by taking
$Z_{Sk}=Z_{Pk}=1$, the flow equation (\ref{flow2}) is further
simplified as
\begin{equation}
\label{flow3}
\partial_k U_k = \frac{k^4}{6\pi^2}T \sum_n \text
{Tr}\left(D_{Sk}+D_{Pk}\right).
\end{equation}

Note that the parameters $h_u$ and $h_s$ (or $h_0$ and $h_8$) which
explicitly break the chiral symmetry are scale independent, and only
the condensates $\bar\sigma_u$ and $\bar\sigma_s$ (or $\bar\sigma_0$
and $\bar\sigma_8$) which spontaneously break the chiral symmetry
depend on the scale $k$. In the treatment of the flow equation with
classical potential, the mass and coupling constant renormalization
leads to four $k$-dependent parameters $m_k, c_k, \lambda_{1k}$ and
$\lambda_{2k}$ controlled by the flow equation (\ref{flow3}), and
the condensates $\bar\sigma_{uk}$ and $\bar\sigma_{sk}$ are
determined by the gap equations (\ref{gap2}).

Considering the partially degenerated diagonal elements and the
simple off-diagonal structure of the mass matrices $M_S$ and $M_P$,
the trace Tr in (\ref{flow3}) can be easily done, and the flow
equation becomes
\begin{eqnarray}
\partial_k U_k &=& \frac{k^4}{6\pi^2}T\sum_n
\big[\text {Tr}\left({\cal D}_{Sk}+{\cal D}_{Pk}\right)\nonumber\\
&+&3D_{a_0k}+4D_{\kappa k}+3D_{\pi k}+4D_{Kk}\big]
\label{flow4}
\end{eqnarray}
with two $2\times 2$ mixed matrices
\begin{eqnarray}
{\cal M}_S &=& \left(\begin{array}{cc}(M_S^2)_{00} & (M_S^2)_{08}\\
(M_S^2)_{80} & (M_S^2)_{88} \end{array}\right),\nonumber\\
{\cal M}_P &=& \left(\begin{array}{cc}(M_P^2)_{00} & (M_P^2)_{08}\\
(M_P^2)_{80} & (M_P^2)_{88} \end{array}\right)
\label{mass3}
\end{eqnarray}
and the corresponding propagators ${\cal
D}^{-1}_{Sk}=\omega_n^2+k^2+{\cal M}_S^2$ and ${\cal
D}^{-1}_{Pk}=\omega_n^2+k^2+{\cal M}_P^2$. Different from the $O(4)$
model\cite{o4frg}, while Tr$(\phi^\dagger\phi)$,
Tr$(\phi^\dagger\phi)^2$ and Det$(\phi)+Det(\phi^\dagger)$ are
invariant under the $SU_L(3)\times SU_R(3)$ transformation, not all
the meson masses can be expressed with the eigenvalues of these
operator composites.

We now expand the potential around the minimum $\bar\sigma_{uk}$ and
$\bar\sigma_{sk}$. Shifting the field
$\sigma_u=\bar\sigma_{uk}+\delta\sigma_{uk}$ and
$\sigma_s=\bar\sigma_{sk}+\delta\sigma_{sk}$, the derivative of the
potential on the left hand side of the flow equation could be
expressed as powers of $\delta\sigma_{uk}$ and $\delta\sigma_{sk}$,
\begin{eqnarray}
\label{expansion}
&& \dot U_k(\bar\sigma_{uk}+\delta\sigma_{uk},\bar\sigma_{sk}+\delta\sigma_{sk})\nonumber\\
&=& \dot U_k+\frac{\partial\dot
U_k}{\partial\delta\sigma_{uk}}\delta\sigma_{uk} +\frac{\partial\dot
U_k}{\partial\delta\sigma_{sk}}\delta\sigma_{sk}+
\frac{1}{2}\frac{\partial^2\dot
U_k}{\partial\delta\sigma_{uk}^2}\delta\sigma^2_{uk}\nonumber\\
&+& \frac{1}{2}\frac{\partial^2\dot
U_k}{\partial\delta\sigma_{sk}^2}\delta\sigma^2_{sk}+
\frac{1}{2}\frac{\partial^2\dot
U_k}{\partial\delta\sigma_{uk}\partial\delta\sigma_{sk}}\delta\sigma_{uk}\delta\sigma_{sk}
\end{eqnarray}
with the definition $\dot U_k=\partial_kU_k$. By comparing the
coefficients of $\delta\sigma_{uk}$, $\delta\sigma_{sk}$,
$\delta\sigma^2_{uk}$, $\delta\sigma^2_{sk}$ and
$\delta\sigma_{uk}\delta\sigma_{sk}$ on the left and right hand
sides of the flow equation (\ref{flow4}), one obtains four
independent differential equations for the six parameters $m^2_k,
c_k, \lambda_{1k}, \lambda_{2k}, \bar\sigma_{uk}$ and
$\bar\sigma_{sk}$. Together with the two gap equations, their
$k$-dependence are fully determined. Note that in deriving the four
flow equations we have used the relations
$\partial_k\delta\sigma_{uk}=-\partial_k\sigma_{uk}$ and
$\partial_k\delta\sigma_{sk}=-\partial_k\sigma_{sk}$.

Before numerically solving the flow equations which will be done in
the next section, we firstly discuss their chiral limit
analytically. From the relation between $h_8$ and $\bar\sigma_8$,
\begin{equation}
\label{kappa}
h_8=m_\kappa^2\bar\sigma_8,
\end{equation}
since there is no reason for the $\kappa$ meson to be massless in
chiral limit, the condensate $\bar\sigma_8$ should vanish in the
case of $h_0=h_8=0$. In this limit, there remains only one gap
equation for the condensate $\bar\sigma_0$,
\begin{equation}
\label{gap3}
m^2-\frac{c}{\sqrt{6}}\bar\sigma_0+\left(\lambda_1
+\frac{\lambda_2}{3}\right)\bar\sigma^2_0=0,
\end{equation}
and the off-diagonal elements in the mass matrices ${\cal M}_{S,P}$
disappear. In the phase of spontaneous chiral symmetry breaking,
there are eight pseudoscalar Goldstone modes $\pi$, K and $\eta$
which dominate the thermodynamics of the system. From the flow
equation (\ref{flow4}), the heavy modes do not contribute much to
the flow, we can keep only the terms with the Goldstone modes. In
this approximation the flow equation for the chiral condensate
$\bar\sigma_{0k}$ is decoupled from the others,
\begin{eqnarray}
\label{flow5}
\partial_k\bar\sigma_{0k}^2 &=& \frac{8k^4}{3\pi^2}T\sum_n\frac{1}{(\omega_n^2+k^2)^2}\nonumber\\
&=& \frac{2k}{3\pi^2}(1+2n_k+2\frac{k}{T}n_k+2\frac{k}{T}n^2_k/T)
\end{eqnarray}
with the Fermi-Dirac distribution $n_k=1/\left(e^{k/T}-1\right)$.

It is easy to see that the right hand side of the flow equation
(\ref{flow5}) is always positive, which leads to a monotonically
increasing $\bar\sigma_{0k}$ with the momentum scale $k$ at any
temperature. Therefore, when we start at an ultraviolet momentum
$\Lambda$, the physical condensate at $k=0$ is guaranteed to be
finite from the evolution of the flow equation. On the other hand,
from the meson frequencies $\omega_n=2n\pi T$, the condensate drops
down with increasing temperature at any momentum scale $k$, which
may lead to a phase transition of chiral symmetry restoration at a
critical temperature $T_c$. In fact, the flow equation can be
analytically integrated out with the solution
\begin{equation}
\label{limit}
\bar\sigma^2_{0k}(T)=\bar\sigma^2_{0\Lambda}(T)+f(\Lambda,T)-f(k,T)
\end{equation}
with the definition
\begin{eqnarray}
\label{fkt} f(k,T) &=&
\frac{T^2}{3\pi^2}\Bigg[5\left(\frac{k}{T}\right)^2+
4\left(\frac{k}{T}\right)^2n_k\nonumber\\
&+&12\frac{k}{T}\text {ln}\left(-n_k\right)-12\text
{Li}_2\left(e^{k/T}\right)\Bigg],
\end{eqnarray}
where Li$_2(x)=\sum_{l=1}^\infty x^l/l^2$ is the polylogarithm
function. At very high momentum, the temperature effect on the
system becomes weak, and we can reasonably take the boundary
condition of the flow equation $\bar\sigma_{0\Lambda}(T)$ at finite
temperature as the one $\bar\sigma_{0\Lambda}(0)$ at zero
temperature. The condensate $\bar\sigma_{0\Lambda}(0)$ is determined
by reproducing the vacuum value $\bar\sigma_{00}(0)$,
\begin{equation}
\label{boundary}
\bar\sigma^2_{0\Lambda}(0)=\bar\sigma^2_{00}(0)+f(0,0)-f(\Lambda,0).
\end{equation}
The critical temperature $T_c$ for the chiral phase transition is
then determined by
\begin{equation}
\label{tc}
\bar\sigma_{0\Lambda}^2(0)+f(\Lambda,T_c)-f(0,T_c)=0.
\end{equation}

In the symmetry restoration phase with $T>T_c$, the potential
expansion should be around the zero condensate $\bar\sigma_0=0$. In
this case, all the mesons become degenerate with mass $m^2$ and
their contributions to the flow equation are the same. The procedure
for the $SU(2)$ model is discussed in \cite{onfrgmeson1,su2frg2}.

\section{Numerical results}
\label{s3} In this section we show our numerical results for chiral
symmetry and $U_A(1)$ symmetry restoration at finite temperature. In
the case with explicit chiral symmetry breaking, one has to solve
the coupled four flow equations, if the wave function
renormalization is neglected, together with the two gap equations.
The initial condition for the four differential equations at a fixed
temperature is the values of the six parameters at the ultraviolet
momentum $\Lambda$, namely $m_\Lambda(T), c_\Lambda(T),
\lambda_{1\Lambda}(T), \lambda_{2\Lambda}(T),
\bar\sigma_{u\Lambda}(T)$ and $\bar\sigma_{s\Lambda}(T)$.
Considering the fact that the system at high enough momentum is
dominated by the dynamics and not affected remarkably by the
temperature, the temperature dependence of the parameters at the
ultraviolet momentum can be safely neglected. Therefore, we take the
temperature independent initial values $m_\Lambda(T)=m_\Lambda(0),
c_\Lambda(T)=c_\Lambda(0),
\lambda_{1\Lambda}(T)=\lambda_{1\Lambda}(0),
\lambda_{2\Lambda}(T)=\lambda_{2\Lambda}(0),
\bar\sigma_{u\Lambda}(T)=\bar\sigma_{u\Lambda}(0)$ and
$\bar\sigma_{s\Lambda}(T)=\bar\sigma_{s\Lambda}(0)$, and they are so
chosen to reproduce their vacuum values at $k=0$ discussed in
Section \ref{s2}, by solving the flow equations and gap equations at
zero temperature.

\begin{figure}[!hbt]
\includegraphics[width=0.5\textwidth]{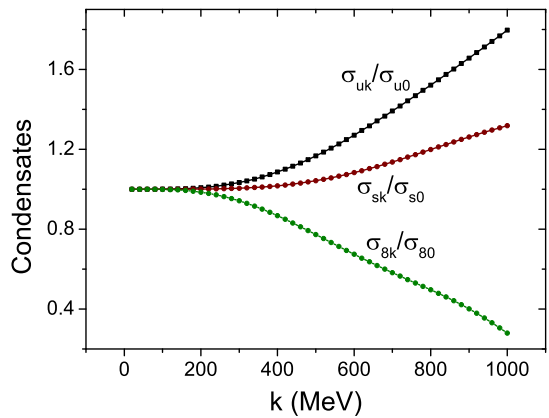}
\caption{(Color online) The evolution of the condensates
$\bar\sigma_{uk}, \bar\sigma_{sk}$ and $\bar\sigma_{8k}$ at $T=0$,
scaled by their vacuum values at $k=0$. }
\label{fig1}
\end{figure}

Fig.\ref{fig1} shows the evolution of the condensates
$\bar\sigma_{uk}, \bar\sigma_{sk}$ and $\bar\sigma_{8k}$ at zero
temperature. With increasing momentum scale $k$, the condensate
$\bar\sigma_{8k}$ drops down continuously and approaches to zero at
$k\sim 1$ GeV. From the relation $h_8=m_{\kappa
k}^2\bar\sigma_{8k}$, the $k$-independence of the chiral breaking
parameter $h_8$ leads to a divergent meson mass $m_{\kappa k}$ at
$k\sim 1$ GeV. Therefore, we take 1 GeV as the ultraviolet momentum
$\Lambda$ for the evolution of the renormalization parameters. We
have checked that the numerical results are insensitive to the value
of the momentum cut $\Lambda$. Both the light and strange quark
condensates $\bar\sigma_{uk}$ and $\bar\sigma_{sk}$ go up with
increasing scale $k$, as we analyzed in the end of Section \ref{s2}
in chiral limit.

\begin{figure}[!hbt]
\includegraphics[width=0.5\textwidth]{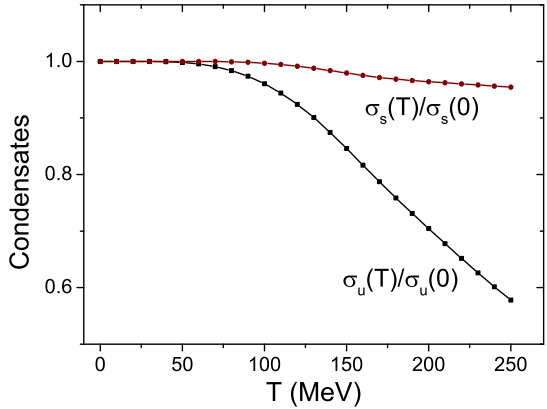}
\caption{(Color online) The temperature dependence of the light and
strange quark condensates $\bar\sigma_{u0}$, $\bar\sigma_{s0}$,
scaled by their vacuum values at $T=0$. }
\label{fig2}
\end{figure}

The temperature dependence of the light and strange quark
condensates at $k=0$ is shown in Fig.\ref{fig2}. While the both
condensates are almost constants at low temperature $T<50$ MeV, they
monotonically decrease at high enough temperature, as we expected
from the analysis in chiral limit. Since strange quarks are much
heavier than light quarks, which is reflected in the values of the
explicit symmetry breaking parameters $h_u$ and $h_s$, the $SU(2)$
symmetry restoration should be much faster than the $SU(3)$ symmetry
restoration. This is the reason why the light quark condensate drops
more rapidly than the strange quark condensate. The qualitative
behave of the condensates shown here is similar to what obtained in
the frame of CJT approximation\cite{cjt}.

\begin{figure}[!hbt]
\includegraphics[width=0.5\textwidth]{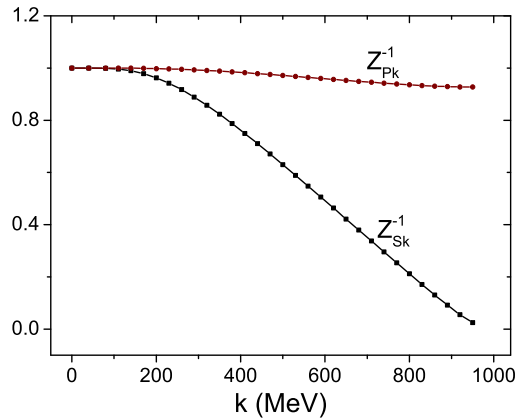}
\caption{(Color online) The momentum scale dependence of the wave
function renormalization constants at zero temperature. }
\label{fig3}
\end{figure}

\begin{figure}[!hbt]
\includegraphics[width=0.5\textwidth]{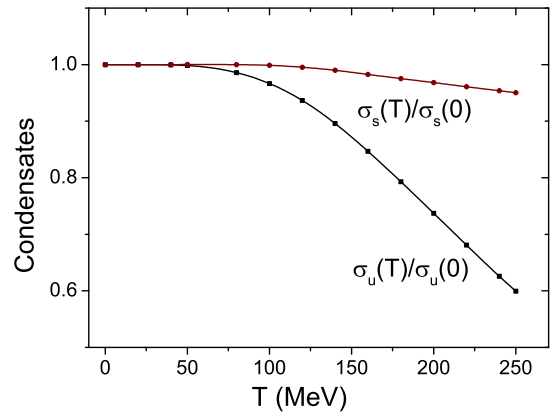}
\caption{(Color online) The temperature dependence of the light and
strange quark condensates, including the wave function
renormalization. }
\label{fig4}
\end{figure}

What is the effect of the wave function renormalization on the
condensates? Solving the flow equations (\ref{expansion}) for the
renormalized mass and coupling constants $m^2_k,\ c_k, \
\lambda_{1k}$ and $\lambda_{2k}$ and (\ref{zk}) for the renormalized
wave functions $Z_{Sk}$ and $Z_{Pk}$ and the gap equations
(\ref{gap2}) for the condensates $\bar\sigma_{uk}$ and
$\bar\sigma_{sk}$, and taking the vacuum values $Z_{S0}=Z_{P0}=1$ at
$T=k=0$, the momentum scale dependence of $Z_{Sk}^{-1}$ and
$Z_{Pk}^{-1}$ at zero temperature are shown in Fig.\ref{fig3}. While
the scalar part varies strongly with the scale $k$, the
renormalization for the pseudoscalar mesons, among which the
Goldstone modes dominate the thermodynamics of the system, is a very
smooth function in the whole region $0<k<\Lambda$. Considering the
fact of $\eta_k\propto \partial_k Z_k^{-1}$ in the flow equation
(\ref{flow2}), the contribution from the wave function
renormalization to the averaged action is expected to be slight.
From Fig.\ref{fig4}, the temperature dependence of the light and
strange quark condensates, including the contribution from the wave
function renormalization, is almost the same as the one shown in
Fig.\ref{fig2}. Therefore, for the following numerical calculations
we will omit the wave function renormalization as a first order
approximation.

The $U_A(1)$ anomaly is described by the topological
susceptibility\cite{lattice1}
\begin{equation}
\label{anomaly}
\chi=\int d^4x\langle 0|T(Q(x)Q(0))|0\rangle
\end{equation}
with $Q(x)$ being the topological charge determined by the QCD
coupling constant $g$ and the gauge field tensor $F^a_{\mu\nu}$,
\begin{equation}
\label{gauge}
Q(x)=\frac{g^2}{64\pi^2}\epsilon^{\mu\nu\rho\sigma}
F^a_{\mu\nu}(x)F^a_{\rho\sigma}(x),
\end{equation}
where $\epsilon_{\mu\nu\rho\sigma}$ is the antisymmetric tensor. By
introducing the anomaly, the ninth Goldstone mode $\eta'$ is no
longer massless. The topological susceptibility can be related to
the pseudoscalar meson masses in the linear sigma model,
\begin{eqnarray}
\label{top} \chi &=& \frac{f^2_\pi}{6}\left(m^2_\eta+m^2_{\eta
'}-2m^2_K\right)\nonumber\\
&=&
\frac{\sigma_u^2}{12}\left(\sqrt{6}c\sigma_0-3\lambda_2\sigma_8^2\right).
\end{eqnarray}
It is determined not only by the three-line coupling $c$ but also by
the four-line coupling $\lambda_2$, although the $U_A(1)$ symmetry
is broken only by the determinant term which is irrelevant to
$\lambda_2$.

\begin{figure}[!hbt]
\includegraphics[width=0.5\textwidth]{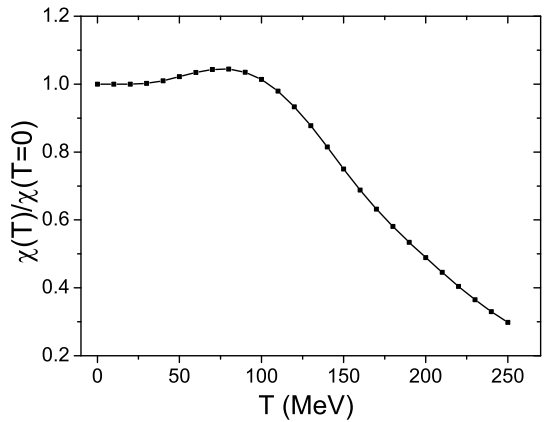}
\caption{(Color online) The temperature dependence of the
topological susceptibility $\chi$, scaled by its vacuum value at
$T=0$. } \label{fig5}
\end{figure}

The temperature dependence of the topological susceptibility is
shown in Fig.\ref{fig5}. It decreases with temperature in the region
of $T\ge 100$ MeV, indicating a continuous restoration of $U_A(1)$
symmetry. Different from the lattice simulation for the $SU(3)$
Yang-Mills theory\cite{lattice1} where the susceptibility drops down
rapidly, the susceptibility remains still $30\%$ of its vacuum value
at high temperature $T=250$ MeV. Unlike the condensates which never
increase with temperature, the susceptibility shows a slight
increase in the low temperature region of $T<100$ MeV, which is
observed also in simulations in the interacting instanton liquid
model\cite{iilm}.

\begin{figure}[!hbt]
\includegraphics[width=0.5\textwidth]{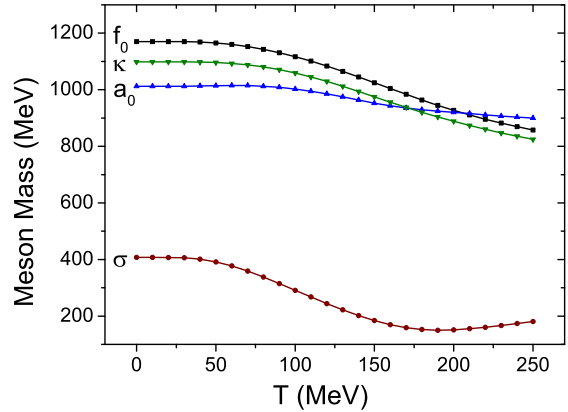}
\caption{(Color online) The temperature dependence of the scalar
meson masses. }
\label{fig6}
\end{figure}

\begin{figure}[!hbt]
\includegraphics[width=0.5\textwidth]{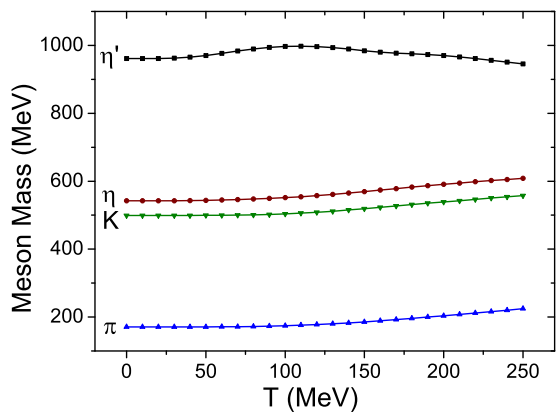}
\caption{(Color online) The temperature dependence of the
pseudoscalar meson masses. }
\label{fig7}
\end{figure}

The chiral symmetry and $U_A(1)$ symmetry restoration shown above
control the meson masses in hot medium. The temperature dependence
of the meson masses is shown in Figs.\ref{fig6} and \ref{fig7}.
There are four different mesons for each species, the triplet $a_0$,
quartet $\kappa$, and the mixed $\sigma$ and $f_0$ for the scalar
mesons, and the triplet $\pi$, quartet K, and the mixed $\eta$ and
$\eta'$ for the pseudoscalar mesons. As temperature increases, the
scalar meson masses $m_{f_0}, m_\kappa$ and $m_{a_0}$ drop down
monotonically, but $f_0$ and $\kappa$ are heavier than $a_0$ at
$T<200$ MeV and become lighter than $a_0$ at $T>200$ MeV. The
Goldstone modes $\pi$ and their chiral partner $\sigma$ become
degenerate at high temperature, due to the chiral symmetry
restoration. The similar behave is for the mesons $a_0$ and $\eta'$.
At $T=250$ MeV there are $m_\pi\simeq m_\sigma\simeq 200$ MeV and
$m_{a_0}\simeq m_{\eta'}\simeq 900$ MeV. The mixed mode $\eta$
behaves similarly to the other seven Goldstone modes $\pi$ and $K$.
While the mass of the other mixed meson $\eta'$ slightly goes up at
low temperature and drops down at high temperature, the difference
between $\eta$ and $\eta'$ decreases with increasing temperature,
due to the partial restoration of the $U_A(1)$ symmetry.

\begin{figure}[!hbt]
\includegraphics[width=0.5\textwidth]{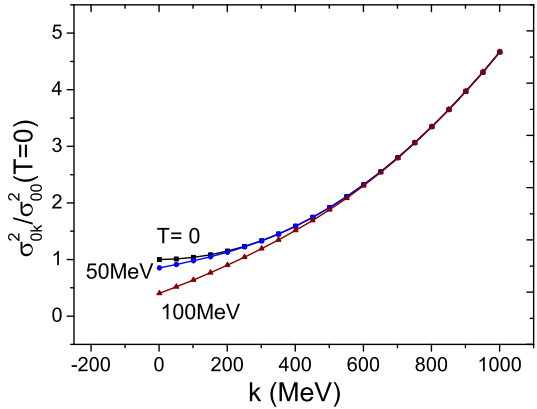}
\caption{The evolutions of the condensate $\bar\sigma_{0k}$ at
different temperature in chiral limit. }
\label{fig8}
\end{figure}

\begin{figure}[!hbt]
\includegraphics[width=0.5\textwidth]{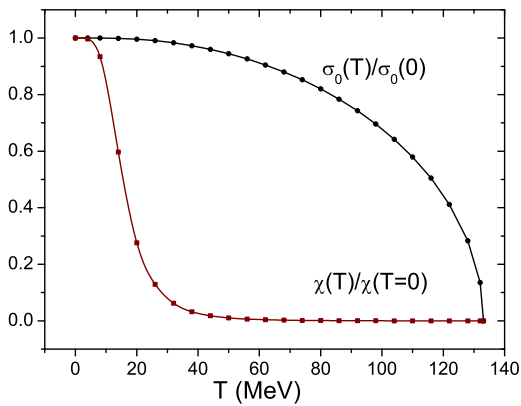}
\caption{The temperature dependence of the chiral condensate
$\bar\sigma_0$ and topological susceptibility $\chi$ in chiral
limit, scaled by their vacuum values.} \label{fig9}
\end{figure}

In chiral limit, the condensate $\bar\sigma_8$ disappears and the
flow equation for the condensate $\bar\sigma_0$ is decoupled from
the others and can be analytically solved as (\ref{limit}). The
evolutions of $\bar\sigma_{0k}$ at different temperature is shown in
Fig.\ref{fig8}. While the temperature dependence at low momentum
scale $k$ is remarkable, the condensate is almost $T$-independent
when the scale is large enough. This supports our choice of
$T$-independent initial condition at $k=\Lambda$. With increasing
temperature, the solution at $k=0$ approaches to zero continuously,
indicating the phase transition of chiral symmetry restoration. The
temperature dependence of the physical condensate and topological
susceptibility at $k=0$ is shown in Fig.\ref{fig9}. The condensate
drops down much faster than that in the general case with explicit
chiral symmetry breaking, see Fig.\ref{fig2}. The critical
temperature for the chiral phase transition determined by
$\bar\sigma_{00}(T_c)=0$ is $T_c=130$ MeV.

The topological susceptibility (\ref{top}) for the $U_A(1)$ anomaly
is simplified as
\begin{equation}
\label{chiral2}
\chi={1\over
3\sqrt 6}c\sigma_0^3
\end{equation}
in chiral limit. After a rapid decrease in the vicinity of vacuum,
it becomes very smooth and finally vanishes at the same critical
temperature for the $SU_A(3)$ symmetry restoration.

\section{Conclusion}
\label{s4} We investigated in this paper the chiral symmetry and
$U_A(1)$ symmetry restoration at finite temperature, by applying the
functional renormalization group to the  $SU(3)$ linear sigma model.
We derived the flow equations for the mass, coupling and wave
function renormalization parameters in the local potential
approximation and the two gap equations for the light and strange
quark condensates.

In chiral limit, we analytically solved the decoupled flow equation
for the chiral condensate and analyzed its momentum scale and
temperature dependence. The eight Goldstone modes are guaranteed in
vacuum and at finite temperature before the chiral restoration, and
the two phase transitions for the chiral symmetry and $U_A(1)$
symmetry restoration take place at the same critical temperature
$T_c=130$ MeV.

In general case with explicit chiral symmetry breaking, we
numerically solved the coupled flow and gap equations at finite
temperature, starting from the classical potential at the
ultraviolet momentum $k=\Lambda\sim 1$ GeV and extracting physics
including quantum fluctuations at $k=0$. In this case a partial
restoration of the $SU_A(3)$ and $U_A(1)$ symmetries is observed.
Different from the results obtained in other approximations like CJT
method, the light and strange quark condensates drop down with
temperature slowly. As a result of the partial restoration of the
two symmetries, the pseudoscalar triplet $\pi$ and its chiral
partner $\sigma$ ( the scalar triplet $a_0$ and $\eta'$) become
degenerate at high temperature, and the difference between the mixed
modes $\eta$ and $\eta'$ gradually disappears in the limit of high
temperature.

\appendix {\bf Acknowledgement:} The work is supported by the NSFC
grant Nos. 10975084 and 11079024.


\begin{thebibliography}{22}
\bibitem{thooft}
  G.~'t Hooft,
  Phys.\ Rev.\  D {\bf 14}, 3432 (1976)
  [Erratum-ibid.\  D {\bf 18}, 2199 (1978)].

\bibitem{eta}
  C.~Rosenzweig, J.~Schechter and C.~G.~Trahern,
  Phys.\ Rev.\  D {\bf 21}, 3388 (1980).

\bibitem{u1restore}
  T.~Schafer,
  Phys.\ Lett.\  B {\bf 389}, 445 (1996)
  [arXiv:hep-ph/9608373].

\bibitem{lattice1}
  B.~Alles, M.~D'Elia and A.~Di Giacomo,
  Nucl.\ Phys.\  B {\bf 494}, 281 (1997)
  [Erratum-ibid.\  B {\bf 679}, 397 (2004)]
  [arXiv:hep-lat/9605013].

\bibitem{lattice2}
  B.~Alles, M.~D'Elia, A.~Di Giacomo and P.~W.~Stephenson,
  Nucl.\ Phys.\ Proc.\ Suppl.\  {\bf 73}, 518 (1999)
  [arXiv:hep-lat/9808004].

\bibitem{iilm}
  O.~Wantz and E.~P.~S.~Shellard,
  Nucl.\ Phys.\  B {\bf 829}, 110 (2010)
  [arXiv:0908.0324 [hep-ph]].

\bibitem{kaon1}
  C.~M.~Ko, Z.~G.~Wu, L.~H.~Xia and G.~E.~Brown,
  Phys.\ Rev.\ Lett.\  {\bf 66}, 2577 (1991)
  [Erratum-ibid.\  {\bf 67}, 1811 (1991)].

\bibitem{kaon2}
  G.~Q.~Li and G.~E.~Brown,
  Phys.\ Rev.\  C {\bf 58}, 1698 (1998)
  [arXiv:nucl-th/9804013].

\bibitem{ratio}
  K.~Paech, A.~Dumitru, J.~Schaffner-Bielich, H.~Stoecker, G.~Zeeb, D.~Zschiesche and S.~Schramm,
  Acta Phys.\ Hung.\  A {\bf 21}, 151 (2004).

\bibitem{u1eta}
  Z.~Huang and X.~N.~Wang,
  Phys.\ Rev.\  D {\bf 53}, 5034 (1996)
  [arXiv:hep-ph/9507395].

\bibitem{u1hydro}
  B.~Keren-Zur and Y.~Oz,
  JHEP {\bf 1006}, 006 (2010)
  [arXiv:1002.0804 [hep-ph]].

\bibitem{bubblelsm}
  G.~Amelino-Camelia,
  Phys.\ Lett.\  B {\bf 407}, 268 (1997)
  [arXiv:hep-ph/9702403].

\bibitem{u1lsm}
  J.~Schaffner-Bielich,
  Phys.\ Rev.\ Lett.\  {\bf 84}, 3261 (2000)
  [arXiv:hep-ph/9906361].

\bibitem{chiral}
  S.~Gasiorowicz and D.~A.~Geffen,
  Rev.\ Mod.\ Phys.\  {\bf 41}, 531 (1969).

\bibitem{mfsu2lsm}
  N.~Bilic and H.~Nikolic,
  Eur.\ Phys.\ J.\  C {\bf 6}, 513 (1999)
  [arXiv:hep-ph/9711513].

\bibitem{su2cjt}
  H.~Mao, N.~Petropoulos and W.~Q.~Zhao,
  J.\ Phys.\ G {\bf 32}, 2187 (2006)
  [arXiv:hep-ph/0606241].

\bibitem{cjt}
  J.~T.~Lenaghan, D.~H.~Rischke and J.~Schaffner-Bielich,
  Phys.\ Rev.\  D {\bf 62}, 085008 (2000)
  [arXiv:nucl-th/0004006].

\bibitem{frgreview1}
  J.~Berges,
  arXiv:hep-ph/9902419.

\bibitem{frgreview2}
  H.~Gies,
  arXiv:hep-ph/0611146.

 \bibitem{frgreview3}
  P.~Kopietz, L.~Bartosch and F.~Schutz,
  Lect.\ Notes Phys.\  {\bf 798}, 1 (2010).

\bibitem{frgcoldatom}
  S.~Floerchinger, R.~Schmidt, S.~Moroz and C.~Wetterich,
  Phys.\ Rev.\  A {\bf 79}, 013603 (2009)
  [arXiv:0809.1675 [cond-mat.supr-con]].

\bibitem{frgnucleon}
  B.~Friman, K.~Hebeler and A.~Schwenk,
  arXiv:1201.2510 [nucl-th].

\bibitem{frgpqmm}
  T.~K.~Herbst, J.~M.~Pawlowski and B.~J.~Schaefer,
  Phys.\ Lett.\  B {\bf 696}, 58 (2011)
  [arXiv:1008.0081 [hep-ph]].

\bibitem{o4frg}
  B.~Stokic, B.~Friman and K.~Redlich,
  Eur.\ Phys.\ J.\  C {\bf 67}, 425 (2010)
  [arXiv:0904.0466 [hep-ph]].

\bibitem{onfrgmeson1}
  O.~Bohr, B.~J.~Schaefer and J.~Wambach,
  Int.\ J.\ Mod.\ Phys.\  A {\bf 16}, 3823 (2001)
  [arXiv:hep-ph/0007098].

\bibitem{onfrgmeson2}
  J.~P.~Blaizot, A.~Ipp, R.~Mendez-Galain and N.~Wschebor,
  Nucl.\ Phys.\  A {\bf 784}, 376 (2007)
  [arXiv:hep-ph/0610004].

\bibitem{su2frg}
  B.~J.~Schaefer and H.~J.~Pirner,
  Nucl.\ Phys.\  A {\bf 660}, 439 (1999)
  [arXiv:nucl-th/9903003].

\bibitem{su2frg2}
  K.~Fukushima, K.~Kamikado and B.~Klein,
  Phys.\ Rev.\  D {\bf 83}, 116005 (2011)
  [arXiv:1010.6226 [hep-ph]].

\bibitem{su3lsm}
  R.~D.~Pisarski and F.~Wilczek,
  Phys.\ Rev.\  D {\bf 29}, 338 (1984).

\bibitem{su3oneloop}
  L.~H.~Chan and R.~W.~Haymaker,
  Phys.\ Rev.\  D {\bf 7}, 402 (1973).

\bibitem{litim}
  D.~F.~Litim,
  Phys.\ Rev.\ D {\bf 64}, 105007 (2001).

\bibitem{regulator}
  G.~Papp, B.~J.~Schaefer, H.~J.~Pirner and J.~Wambach,
  Phys.\ Rev.\ D {\bf 61}, 096002 (2000).
\end{thebibliography}
\end{document}